\documentclass[%
 aip,
 sd,%
 amsmath,amssymb,
preprint,%
]{revtex4-1}

\usepackage{graphicx}
\usepackage{dcolumn}
\usepackage{bm}
\usepackage[english]{babel}
\usepackage{amsmath,amsthm}
\usepackage{amsfonts}
\usepackage{xspace}
\usepackage{bm}
\usepackage{booktabs}
\usepackage{graphicx}
\usepackage{dcolumn}
\usepackage[mathlines]{lineno}
\usepackage[colorlinks]{hyperref}  
\usepackage{gensymb}
\usepackage{epstopdf}
\usepackage{newfloat}

\begin{document}


\title{Control of magnetic relaxation by electric-field-induced ferroelectric phase transition and inhomogeneous domain switching}

\author{\textit{Tianxiang Nan}}
 \affiliation{Department of Electrical and Computer Engineering, Northeastern University, Boston, Massachusetts 02115,USA}
\author{\textit{Satoru Emori}}
\affiliation{Department of Electrical and Computer Engineering, Northeastern University, Boston, Massachusetts 02115,USA}%
\author{\textit{Bin Peng}}
 \affiliation{Electronic Materials Research Laboratory, Xi'an Jiaotong University, Xi'an 710049, China}%
 \author{\textit{Xinjun Wang}}
 \affiliation{Department of Electrical and Computer Engineering, Northeastern University, Boston, Massachusetts 02115,USA}%
 \author{\textit{Zhongqiang Hu}}
 \affiliation{Department of Electrical and Computer Engineering, Northeastern University, Boston, Massachusetts 02115,USA}%
 \author{\textit{Li Xie}}
 \affiliation{Department of Electrical and Computer Engineering, Northeastern University, Boston, Massachusetts 02115,USA}%
 \author{Yuan Gao}
 \affiliation{Department of Electrical and Computer Engineering, Northeastern University, Boston, Massachusetts 02115,USA}%
  \author{Hwaider Lin}
 \affiliation{Department of Electrical and Computer Engineering, Northeastern University, Boston, Massachusetts 02115,USA}%
 \author{Jie Jiao}
 \affiliation{Shanghai Institute of Ceramics, Chinese Academy of Sciences, Shanghai 201800, China}%
 \author{Haosu Luo}
 \affiliation{Shanghai Institute of Ceramics, Chinese Academy of Sciences, Shanghai 201800, China}%
 \author{David Budil}
 \affiliation{Department of Chemistry, Northeastern University, Boston, Massachusetts 02115, USA }%
 \author{John G. Jones}
 \affiliation{Materials and Manufacturing Directorate, Air Force Research Laboratory, Wright-Patterson AFB, Ohio 45433, USA}%
 \author{Brandon M. Howe}
 \affiliation{Materials and Manufacturing Directorate, Air Force Research Laboratory, Wright-Patterson AFB, Ohio 45433, USA}%
  \author{Gail J. Brown}
 \affiliation{Materials and Manufacturing Directorate, Air Force Research Laboratory, Wright-Patterson AFB, Ohio 45433, USA}%
 \author{Ming Liu}
  \email{mingliu@mail.xjtu.edu.cn}
 \affiliation{Electronic Materials Research Laboratory, Xi'an Jiaotong University, Xi'an 710049, China}
  \author{Nian Sun}
  \email{n.sun@neu.edu}
 \affiliation{Department of Electrical and Computer Engineering, Northeastern University, Boston, Massachusetts 02115,USA}%

\date{\today}

\begin{abstract}
Electric-field modulation of magnetism in strain-mediated multiferroic heterostructures is considered a promising scheme for enabling memory and magnetic microwave  devices with ultralow power consumption. 
However, it is not well understood how electric-field-induced strain influences magnetic relaxation, an important physical process for device applications.
Here we investigate resonant magnetization dynamics in ferromagnet/ferrolectric multiferroic heterostructures, FeGaB/PMN-PT and NiFe/PMN-PT, in two distinct strain states provided by electric-field-induced ferroelectric phase transition. 
The strain not only modifies magnetic anisotropy but also magnetic relaxation.
In FeGaB/PMN-PT, we observe a nearly two-fold change in intrinsic Gilbert damping by electric field, which is attributed to strain-induced tuning of spin-orbit coupling.
By contrast, a small but measurable change in extrinsic linewidth broadening is attributed to inhomogeneous ferroelastic domain switching during the phase transition of the PMN-PT substrate.
\end{abstract}

\maketitle

%


Electrical manipulation of the magnetization state is essential for improving the scalability and power efficiency of magnetic random access memory (MRAM)~\cite{Chappert2007,Brataas2012,Shiota2012,Hu2011a,Bauer2014}. A particularly promising scheme relies on an electric field to assist or induce magnetization switching with minimal power dissipation\cite{Bibes2008,Hu2011a,Tsymbal2012}. 
Multiferroic magnetoelectric materials with coupled magnetization and electric polarization offer possibilities for electric-field-driven magnetization switching at room temperature~\cite{Eerenstein2006,Fiebig2005,Ramesh2007,Ma2011,Sun2012,Vaz2012,Fusil2014}. Such magnetoelectric effects have been demonstrated with strain-~\cite{Nan1994,Srinivasan2002,Thiele2007,Jia2008a,Lou2009a}, charge-~\cite{Weisheit2007,Maruyama2009,Wang2012,Zhou2013} and exchange bias mediated coupling mechanisms~\cite{Baek2010,Heron2011,Heron2014,Zhou2015}. For example, non-volatile magnetization switching with remarkable modulation of magnetic anisotropy was realized using electric-field-induced piezo-strain at the interface between ferromagnetic and ferroelectric phases~\cite{Wu2011c,Zhang2012,Nan2012a,Liu2013a}. 

On the other hand, a better understanding of the processes responsible for magnetic relaxation, especially at various strain states, is required for electric-field-assisted MRAM or tunable magnetic microwave devices. Recent studies suggest that electric-field-induced changes of magnetic relaxation are correlated to the piezo-strain state or effective magnetic anisotropy~\cite{Lou2008,Zhou2014a,Zighem2014,Li2014a,Yu2015}. A similar modulation of magnetic relaxation has also been observed in a charge-mediated magnetoelectric heterostructure with ultra-thin ferromagnets~\cite{Okada2014}.
In general, the contributions to magnetic relaxation include intrinsic Gilbert damping due to spin-orbit coupling and extrinsic linewidth broadening due to inhomogeneity in the ferromagnet. So far, the understanding of how a piezo-strain modifies these intrinsic and extrinsic contributions has been limited~\cite{Zhou2014a}.

In this work, we quantify electric-field-induced modifications of both intrinsic Gilbert damping and inhomogeneous linewidth broadening in two ferromagnet/ferroelectric hetereostructures: Fe$_{7}$Ga$_{2}$B$_{1}$/Pb(Mg$_{1/3}$Nb$_{2/3}$)O$_{3}$-PbTiO$_{3}$ (FeGaB/PMN-PT) with a strong strain-mediated magnetoelectric (magnetostrictive) coupling and Ni$_{80}$Fe$_{20}$/Pb(Mg$_{1/3}$Nb$_{2/3}$)O$_{3}$-PbTiO$_{3}$ with a negligible magnetoelectric coupling.
The rhombohedral (011) oriented PMN-PT substrate provides two distinct strain states through an electric-field-induced phase transformation~\cite{Shanthi2009,Li2011}. We conduct ferromagnetic resonance (FMR) measurements at several applied electric field values to disentangle the intrinsic and extrinsic contributions to magnetic relaxation. FeGaB/PMN-PT exhibits pronounced electric-field-induced modifications of the resonance field and intrinsic Gilbert damping, whereas these parameters remain mostly unchanged for NiFe/PMN-PT. These findings show that magnetic relaxation can be tuned through a strain-mediated modification of spin-orbit coupling in a highly magnetostrictive ferromagnet.  We also observe in both multiferroic hetereostructures a small electric-field-induced change in extrinsic linewidth broadening, which we attribute to the ferroelectric domain state in the PMN-PT substrate.  

30-nm thick films of FeGaB and NiFe were sputter-deposited on (011) oriented PMN-PT single crystal substrates buffered with 5-nm thick Ta seed layers. The FeGaB thin film was co-sputtered from Fe$_{80}$Ga$_{20}$ (DC sputtered) and B (RF sputtered) targets. Both FeGaB and NiFe films were capped with 2 nm of Al to prevent oxidation. All films were deposited in 3 mTorr Ar atmosphere with a base pressure $\le$ $1\times 10^{-7}$ Torr. The thicknesses of deposited films were calibrated by X-ray reflectivity.

\begin{figure}
\includegraphics[width=0.7\columnwidth]{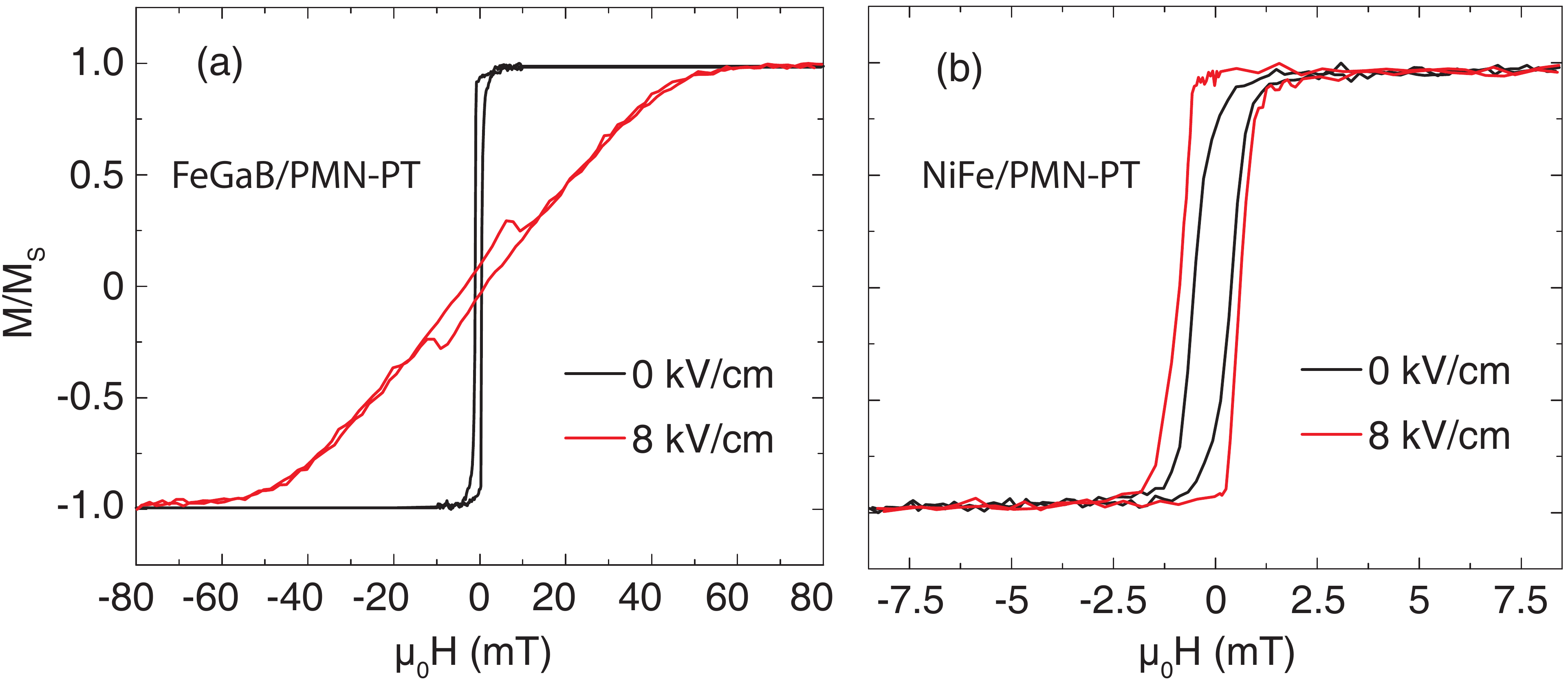}
\caption{(a,b) Electric-field dependent magnetic hysteresis loops with the magnetic field applied along the [100] direction for FeGaB/PMN-PT (a) and NiFe/PMN-PT (b).} 
\label{fig1}
\end{figure}

The amorphous FeGaB thin film was selected for its high saturation magnetostriction coefficient of up to 70 ppm~\cite{Lou2007} and large magnetoelectric effect when interfaced with ferroelectric materials~\cite{Lou2009a}. NiFe was chosen as the control sample with near zero magnetostriction; the thickness of 30 nm is far above the thickness regime that shows high surface magnetostricion~\cite{Song1994}. Fig.~\ref{fig1} shows magnetic hysteresis loops of FeGaB/PMN-PT and NiFe/PMN-PT, measured by vibrating sample magnetometry with an in-plane magnetic field applied along the [100] direction of PMN-PT. An electric field was applied in the thickness direction of the PMN-PT substrate. Due to the anisotropic piezeoelectric coefficient of PMN-PT, an in-plane compressive strain is induced along the [100] direction, which results in uniaxial magnetic anisotropy along the same axis. In FeGaB/PMN-PT the electric field ($E$ = 8 kV/cm) increases the saturation field by $\approx$40 mT, whereas only a small change is observed in NiFe/PMN-PT, confirming the significantly different strengths of strain-mediated magnetoelectric coupling for the two multiferroic heterostructures.

\begin{figure}
\includegraphics[width=0.7\columnwidth]{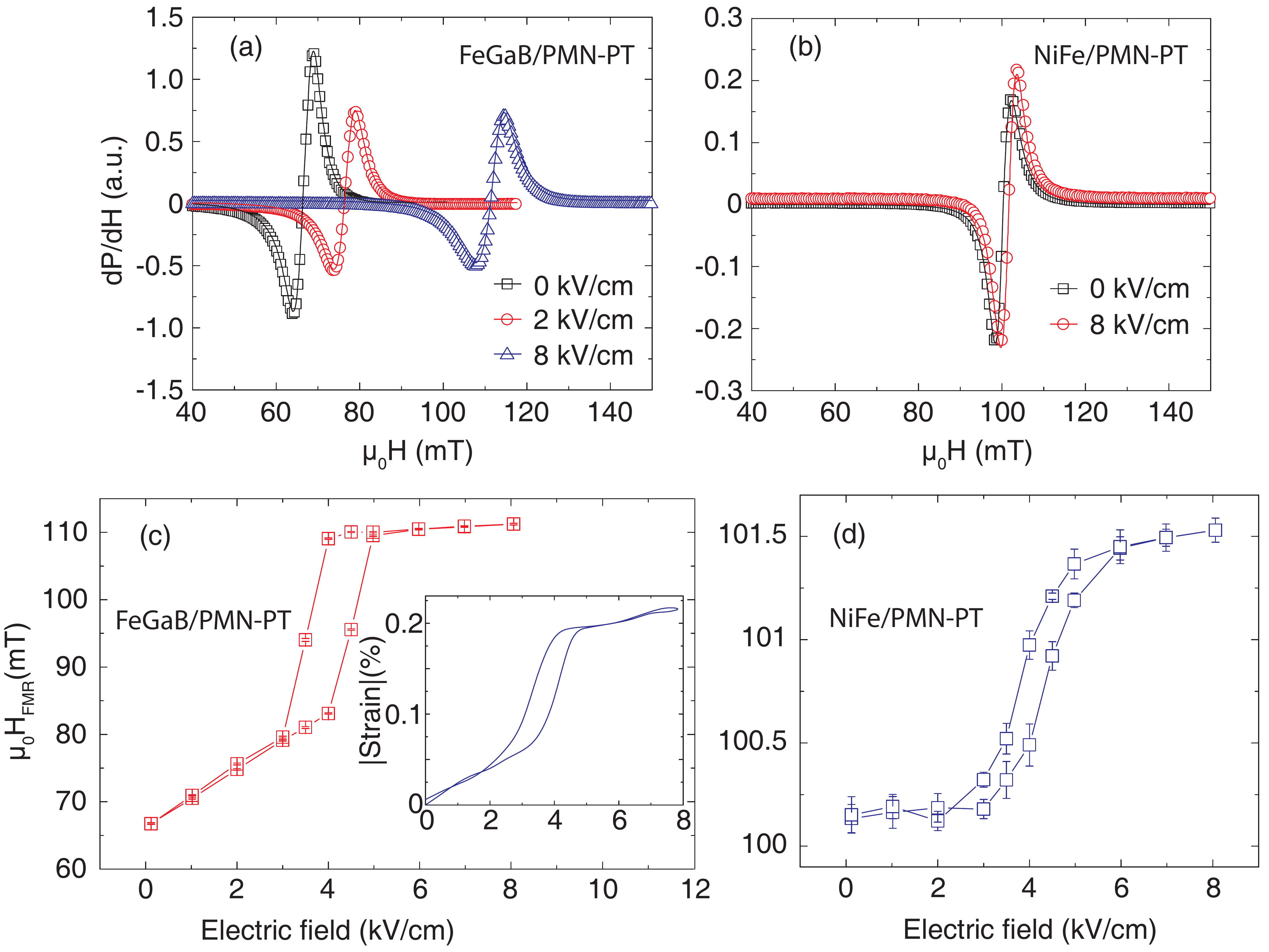}
\caption{(a,b) FMR (fixed at 9.5 GHz) spectra at various electric fields with the magnetic field applied along the [100] direction for FeGaB/PMN-PT (a) and NiFe/PMN-PT (b). (c,d) Resonance field HFMR as a function of the applied electric field for FeGaB/PMN-PT (a) and NiFe/PMN-PT (b). Inset of (c) shows the piezo-strain as a function of electric field for PMN-PT substrate along the [100] direction.} 
\label{fig2}
\end{figure}

Both ferromagnetic thin films exhibit comparatively narrow resonant linewidths, allowing for sensitive detection of the electric-field modification of spin relaxation.  Electric-field dependent FMR spectra of FeGaB/PMN-PT and NiFe/PMN-PT were measured using a Bruker EMX electron paramagnetic resonance (EPR) spectrometer with a TE$_{102}$ cavity operated at a microwave frequency of 9.5 GHz.  The external magnetic field was applied along the [100] direction of the PMN-PT single crystal. These spectra, shown in Fig.~\ref{fig2}(a), (b) were fitted to the derivative of a modified Lorentzian function~\cite{Stancik2008} to extract the resonance field $H_{FMR}$ and resonance linewidth $W$. In FeGaB/PMN-PT, upon applying $E$ = 2 kV/cm along the thickness direction of PMN-PT, a slight increase of $H_{FMR}$ by 10 mT is observed. A larger shift of 35 mT in $H_{FMR}$ is induced at $E$ = 8 kV/cm. In comparison, NiFe/PMN-PT exhibits a much smaller $H_{FMR}$ shift of 1.5 mT at $E$ = 8 kV/cm, as shown in Fig.~\ref{fig2}(b).

The shift of $H_{FMR}$ in FeGaB/PMN-PT and NiFe/PMN-PT as a function of E is summarized in Fig.~\ref{fig2}(c) and (d). Both samples show hysteric behavior that follows the piezo-strain curve of PMN-PT (inset of Fig.~\ref{fig2}(c)) measured with a photonic sensor. This can be understood by the strain-mediated magnetoelectric coupling with the electric-field-induced change of magnetic anisotropy field ($\Delta$H$_{k}$) expressed by
\begin{equation}\label{eq:dHk}
\begin{split}
\Delta H_{k} = \frac{3\lambda(\sigma_{100}-\sigma_{0-11}) }{\mu_{0}M_{s}}
\end{split}
\end{equation}
where $\sigma_{100}$ and $\sigma_{0-11}$ are the in-plane piezo-stress, $\lambda$ and $M_{s}$ are the magnetostriction constant and the saturation magnetization respectively. Considering an in-plane compressive strain along the [100] direction and a positive magnetostriction coefficient of both FeGaB and NiFe, a decrease of the magnetic anisotropy field $H_{k}$ is expected with a positive electric field. The drop of $H_{k}$ results in an increase of $H_{FMR}$ described by the Kittel equation,
\begin{equation}\label{eq:kittel}
f /2\pi= \gamma\mu_{0}\sqrt{(H_{FMR}+H_{k})(H_{FMR}+H_{k}+M_{eff})}
\end{equation}
where $\gamma/2\pi$=28 GHz/T and $M_{eff}$is the effective magnetization. At $E$ ${<}$ 4 kV/cm, $H_{FMR}$ increases linearly, which corresponds to the linear region of piezoelectric effect of PMN-PT with a uniaxial compressive piezo-strain along [100] direction. The sudden change of $H_{FMR}$ at $E$ = 4 kV/cm is attributed to the rhombohedral-to-orthorhombic (R-O) phase transition of PMN-PT substrate~\cite{Li2011}. The PMN-PT substrate reverts to the rhombohedral phase upon decreasing the electric field. Therefore, the R-O phase transformation with a large uniaxial in-plane strain induces two stable and reversible magnetic states at $E$ = 0 and 8 kV/cm. This provides a reliable platform for studying magnetization dynamics in a controlled manner with the applied electric field.

\begin{figure}
\includegraphics[width=0.7\columnwidth]{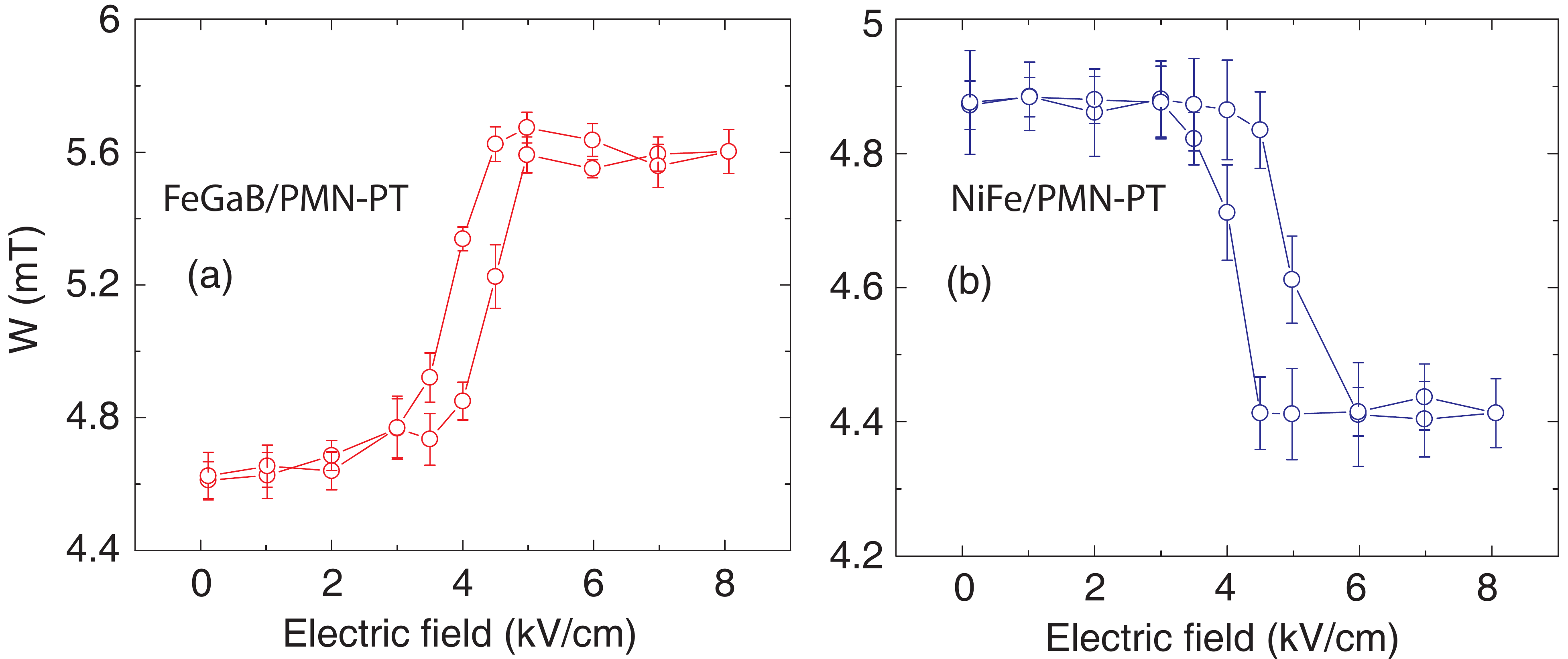}
\caption{(a,b) Resonance linewidth W at 9.5 GHz with the magnetic field applied along the [100] direction as a function of the applied electric field for FeGaB/PMN-PT (a) and NiFe/PMN-PT (b).} 
\label{fig3}
\end{figure}

The peak-to-peak FMR linewidth $W$ of FeGaB/PMN-PT and NiFe/PMN-PT, extracted from the same FMR measurements in Fig.~\ref{fig2}, also exhibits a strong dependence on the applied electric field as shown in Fig.~\ref{fig3}. For FeGaB/PMN-PT, $W$ remains unchanged within experimental uncertainty at $E$ ${<}$4 kV/cm and abruptly increases from $\approx$4.6 mT to $\approx$5.6 mT across the R-O phase transition. By removing the applied electric field, $W$ decreases to the original value with the reversal to the rhombohedral phase. Comparing Fig.~\ref{fig2}(c) and~\ref{fig3}(a), it is evident that the observed electric-field-induced changes in $H_{FMR}$ and $W$ in FeGaB/PMN-PT are correlated, consistent with recent studies~\cite{Zighem2014,Li2014a,Yu2015}.  The change in $W$ indicates a modulation in spin-orbit coupling in the ferromagnet; considering that spin-orbit coupling governs the intrinsic Gilbert damping, it is reasonable that we observe simultaneous modification of $W$ and $H_{FMR}$ by strain in the magnetostrictive FeGaB film.

Given the same sign of the magnetostriction coefficient for FeGaB and NiFe~\cite{Lou2007,Bonin2005}, one would expect to also observe a small increase in $W$ with increasing electric field across the R-O phase transition in NiFe/PMN-PT. However, NiFe/PMN-PT exhibits a decrease in $W$ across the phase transition. This observation indicates that the piezo-strain modifies a different magnetic relaxation contribution in NiFe.

\begin{figure}
\includegraphics[width=0.7\columnwidth]{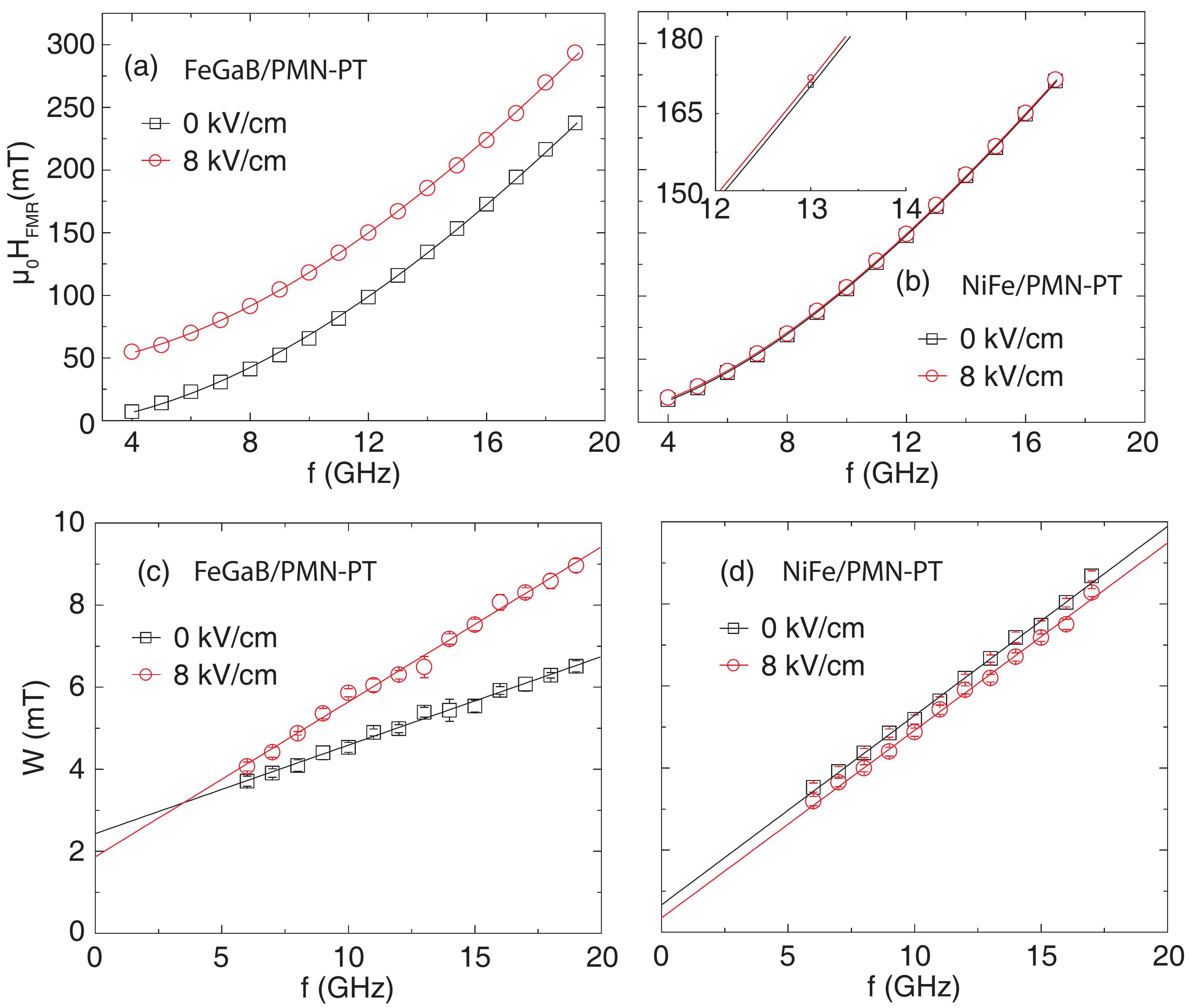}
\caption{(a,b) Frequency f as a function of resonance field HFMR at different electric fields for FeGaB/PMN-PT (a) and NiFe/PMN-PT (b). (c,d) Linewidth W as a function of frequency f at different electric fields for FeGaB/PMN-PT (c) and NiFe/PMN-PT (d). The magnetic field was applied along the [100] direction.} 
\label{fig4}
\end{figure}

The FMR linewidth $W$ consists of the intrinsic Gilbert damping contribution (parameterized by the damping constant $\alpha$) and the frequency-independent inhomogeneous linewidth broadening $W_{0}$: 
\begin{equation}\label{eq:damping}
W=W_{0}+\frac{4\pi\alpha}{\sqrt{3}\gamma}f
\end{equation}
where $f$ is the microwave excitation frequency.  According to Eq.~\ref {eq:damping} , $\alpha$ and $W_{0}$ can be determined simply by measuring the frequency dependence of $W$.  For this purpose, we used a home-built broadband FMR system~\cite{Beguhn2012} with a nominal microwave power of -5 dBm and $f$ = 6-19 GHz. Just as in the single-frequency measurement using the EPR system (Fig.~\ref{fig2} and Fig.~\ref{fig3}), the external magnetic field was applied along the [100] direction of the PMN-PT substrate. By fitting the frequency dependence of $H_{FMR}$ to Eq.~\ref {eq:dHk} (Fig.~\ref{fig4}(a), (b)), we obtain anisotropy field shift $\Delta H_{k} \approx$ 46 mT for FeGaB/PMN-PT and $\Delta H_{k} \approx$ 1 mT for NiFe/PMN-PT across the R-O phase transition, in agreement with the single-frequency FMR measurement (Fig.~\ref{fig2}), while $\mu_{0}M_{eff}$  remains unchanged.  Fig.~\ref{fig4}(c) and (d) plot $W$ as a function of the frequency for FeGaB/PMN-PT and NiFe/PMN-PT, respectively. From the slope of the linear fit (Eq.~\ref {eq:damping}), we find that $\alpha$ of FeGaB/PMN-PT increases from $(0.6\pm0.01)\times10^{-2}$ at $E$ = 0 to $(1.06\pm0.02)\times10^{-2}$ at $E$ = 8 kV/cm, whereas $\alpha$ is unchanged at $(1.29\pm0.16)\times10^{-2}$ for NiFe/PMN-PT within experimental uncertainty($\alpha$=$(1.27\pm0.2)\times10^{-2}$ at $E$ = 8 kV/cm). The large change in $\alpha$ for FeGaB and negligible change for NiFe suggest a strong correlation between magnetostriction and the intrinsic Gilbert damping mechanism. In particular, a large in-plane uniaxial strain generated by the R-O phase transformation induces an additional anisotropy field in FeGaB that enhances the dephasing of the magnetization precession~\cite{Bonin2005}.  

\begin{figure}
\includegraphics[width=0.7\columnwidth]{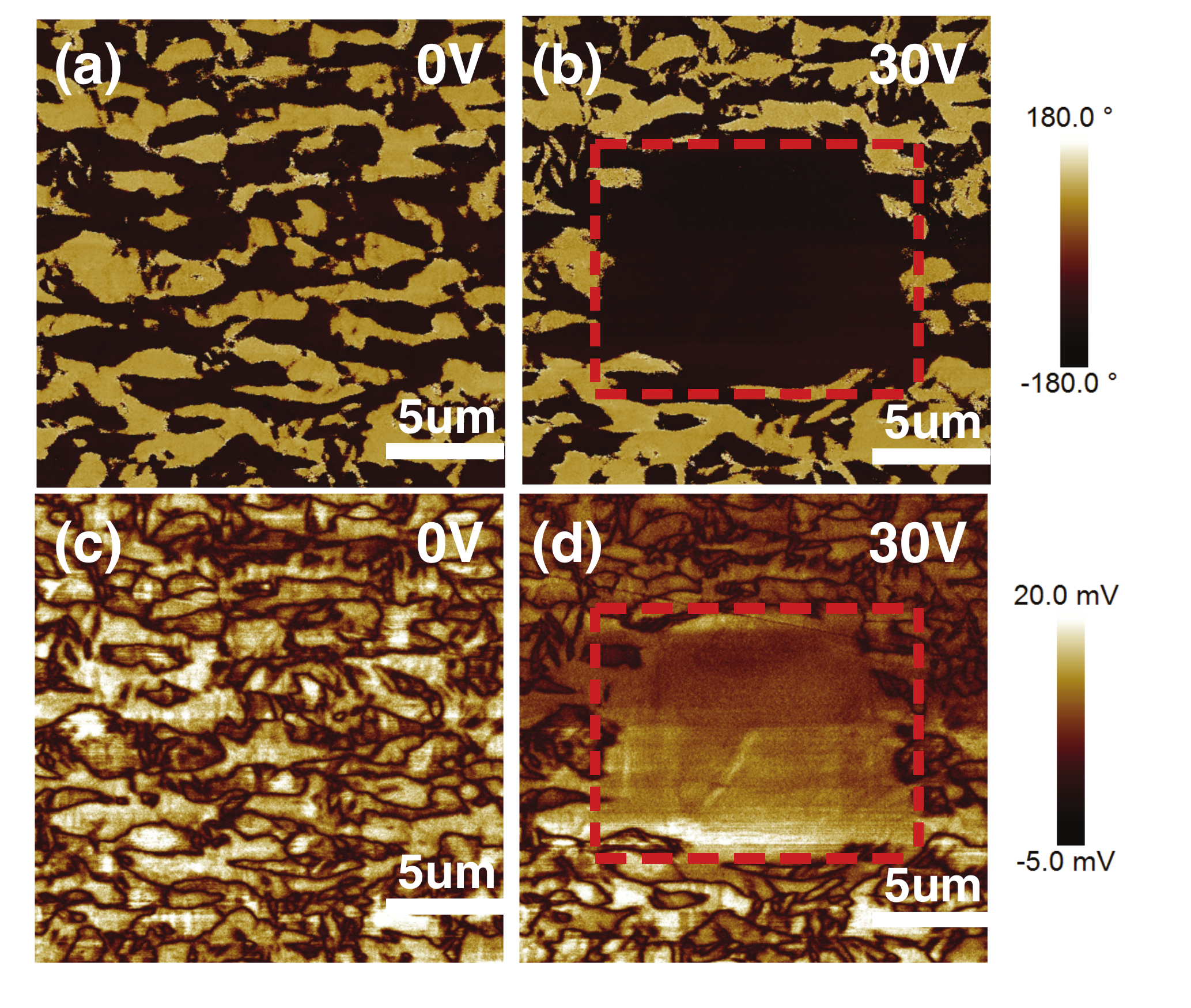}
\caption{(a,b) The out-of-plane vertical PFM (VPFM) phase images upon applying different voltages to the square area outlined by a red dashed line. (c,d) Corresponding amplitude images at different voltage biases.} 
\label{fig5}
\end{figure}

However, both FeGaB/PMN-PT and NiFe/PMN-PT show a decreased $W_{0}$ upon applying $E$ = 8 kV/cm. This could be related to the ferroelectric domain state in the PMN-PT substrate that significantly affects the homogeneity of the magnetic film on top. The polarization domain phase images with various applied voltages are shown in Fig.~\ref{fig5} by using a piezo-force microscope. For the unpoled state at, as shown in Fig.~\ref{fig5} (a), the polarization state of PMN-PT surface is inhomogenous, with polarization vectors oriented randomly along the eight body diagonals of the pseudocubic cell. By applying a voltage of 30 $V$ within the gated area (dashed outline in Fig.~\ref{fig5} (b),(d)), the ferroelectric state becomes saturated within this area with all the polarization vectors pointing upward. This uniformly polarized state alters the surface topology the PMN-PT substrate~\cite{Liu2013a}, thereby reducing the inhomogeneous linewidth broadening $W_{0}$ of the ferromagnetic film.  

We also measured frequency-dependent FMR spectra with an external magnetic field applied along the [0$\bar{1}$1] direction to examine the anisotropy of magnetic relaxation. For FeGaB/PMN-PT, $\alpha$ and $W_{0}$ are close to the [100] configuration at $E$ = 0. At $E$ = 8 kV/cm, we observed a non-linear relation between $W$ and $f$, which might have resulted from a highly non-uniform magnetization state at low fields due to the large electric-field-induced $H_{k}$~\cite{Woltersdorf2004,Zakeri2007}. To extract $\alpha$ reliably in this case, we would need to conduct FMR measurements at higher frequencies. For NiFe/PMN-PT, $\alpha$ and the electric-field dependence of $W_{0}$ are identical for the [0$\bar{1}$1] and [100] directions. The parameters quantified in this study are summarized in Table~\ref{tab:params}.  

In summary, we have quantified electric-field-induced modifications of magnetic anisotropy and magnetic relaxation contributions, namely intrinsic Gilbert damping and inhomogeneous linewidth broadening, in multiferroic heterostructures. A large modification of intrinsic damping arises from strain-induced tuning of spin-orbit coupling in the ferromagnet and is correlated with the magnitude of magnetostriction. A small change in the extrinsic linewidth contribution is attained by controlling the ferroelectric domain states in the substrate. These findings are not only of technology importance for the application on low-power MRAM and magnetic microwave devices, but also permit investigation of the structural dependence of spin-orbit-derived phenomena in magnetic thin films.\\

\begin{center}
\begin{table}
\caption{Parameters extracted from broadband FMR at 2 different electric fields} 
\centering 
\begin{tabular}{ccccc} 
\hline\hline 
{}&\multicolumn{2}{c}{FeGaB/PMN-PT}&\multicolumn{2}{c}{NiFe/PMN-PT} \\ [0.5ex] 
\hline 
{$E$(kV/cm)}&\multicolumn{1}{c}{0}&\multicolumn{1}{c}{8}&\multicolumn{1}{c}{0}&\multicolumn{1}{c}{8}  \\ [0.5ex] 
\hline 
$4\pi$$M_{eff}(T)$ &\multicolumn{1}{c}{$1.48\pm0.01$}&\multicolumn{1}{c}{$1.46\pm0.01$}&\multicolumn{1}{c}{$0.96\pm0.04$}&\multicolumn{1}{c}{$0.96\pm0.04$}  \\ 
$H_{k}(mT)[100]$ &\multicolumn{1}{c}{$5.8\pm0.5$}&\multicolumn{1}{c}{$-41.3\pm0.3$}&\multicolumn{1}{c}{$1.67\pm0.2$}&\multicolumn{1}{c}{$0.27\pm0.2$}  \\ 
$\alpha(10^{-2})[100]$ &\multicolumn{1}{c}{$0.6\pm0.01$}&\multicolumn{1}{c}{$1.06\pm0.02$}&\multicolumn{1}{c}{$1.29\pm0.16$}&\multicolumn{1}{c}{$1.27\pm0.2$}  \\ 
$W_{0}(mT)[100]$ &\multicolumn{1}{c}{$2.4\pm0.05$}&\multicolumn{1}{c}{$1.8\pm0.07$}&\multicolumn{1}{c}{$0.66\pm0.06$}&\multicolumn{1}{c}{$0.35\pm0.07$}  \\ 
$H_{k}(mT)[0\bar{1}1]$ &\multicolumn{1}{c}{$3.24\pm0.4$}&\multicolumn{1}{c}{$\footnote{Not able to obtain due to the frequency constraint and the field-dragging effect at measured low frequencies.}$}&\multicolumn{1}{c}{$1.54\pm0.2$}&\multicolumn{1}{c}{$3.1\pm0.3$}  \\ 
$\alpha(10^{-2})[0\bar{1}1]$ &\multicolumn{1}{c}{$0.6\pm0.02$}&\multicolumn{1}{c}{}&\multicolumn{1}{c}{$1.21\pm0.12$}&\multicolumn{1}{c}{$1.29\pm0.15$}  \\ 
$W_{0}(mT)[0\bar{1}1]$ &\multicolumn{1}{c}{$2.8\pm0.05$}&\multicolumn{1}{c}{}&\multicolumn{1}{c}{$5.9\pm0.08$}&\multicolumn{1}{c}{$2.9\pm0.05$}  \\ 

\hline 
\end{tabular}
\label{tab:params}
\end{table}
\end{center}

\newpage

This work was supported by the Air Force Research Laboratory through contract FA8650-14-C-5706 and in part by FA8650-14-C-5705, the W.M. Keck Foundation, and the National Natural Science Foundation of China (NSFC) 51328203, 51472199, 11534015.

\newpage

\bibliography{library1}
\bibliographystyle{aipnum4-1}

\newpage

\end{document}